\documentclass[conference]{IEEEtran}
\IEEEoverridecommandlockouts
\usepackage{cite}
\usepackage{amsmath,amssymb,amsfonts}
\usepackage{supertabular,booktabs}
\usepackage{graphicx}
\usepackage{textcomp}
\usepackage{xcolor}
\usepackage{bbm}
\usepackage{nicefrac, xfrac}
\usepackage{mathtools}
\usepackage{algpseudocode}
\usepackage{enumitem}
\DeclareMathOperator*{\argmin}{arg\,min}

\def\BibTeX{{\rm B\kern-.05em{\sc i\kern-.025em b}\kern-.08em
    T\kern-.1667em\lower.7ex\hbox{E}\kern-.125emX}}

\newcommand{\cR}{\mathcal{R}}
\newcommand{\bcR}{\bar{\mathcal{R}}}
\newcommand{\bw}{\mathbf{w}}
\newcommand{\bh}{\mathbf{h}}

\newtheorem{remark}{Remark}

\begin{document}

\title{Least-Cost Structuring of 24/7 Carbon-Free Electricity Procurements
\thanks{Work partially supported by ARPA-E PERFORM grant DE-AR0001289.}
}

\author{\IEEEauthorblockN{Mike Ludkovski, Saad Mouti}
\IEEEauthorblockA{\textit{Dept of Statistics \& Applied Probability} \\
\textit{UC Santa Barbara}\\
Santa Barbara, CA USA \\
\{ludkovski,smouti\}@pstat.ucsb.edu}
\and
\IEEEauthorblockN{Glen Swindle}
\IEEEauthorblockA{\textit{Scoville Risk Partners} \\
Princeton, NJ USA}}

\date{October 2023}

\maketitle

\begin{abstract}
We consider the construction of renewable portfolios targeting specified carbon-free (CFE) hourly performance scores. We work in a probabilistic framework that uses a collection of simulation scenarios and imposes probability constraints on achieving the desired CFE score. In our approach there is a fixed set of available CFE generators and a given load customer who seeks to minimize annual procurement costs. We illustrate results using a realistic dataset of jointly calibrated solar and wind assets, and compare different approaches to handling multiple loads.
\end{abstract}


\section{Introduction}
We consider the problem of matching a given load profile with a portfolio of renewable generators to achieve a desired 24/7 CFE score. This has become a topical task for many large industrial users, especially data centers and tech companies, who seek auditable (via e.g.~ISO settlement hourly reports or renewable energy credit accounts) commitment to reducing their carbon footprint. Corporate customers with the appetite to ``do the right thing"  procure carbon-free energy sources on a purchasing power agreement (PPA) basis and evaluate their hourly CFE metric, i.e.~the ratio of total load to purchased CFE. Crucially, unused CFE does not contribute, while unmatched load is penalized linearly. The goal is then to hit a desired \emph{annual} target.  Due to inherent intermittency, such hourly matching is an essential but technically challenging step to achieving 100\% decarbonization \cite{miller2022hourly,xu2022electricity}. 

Our work is inspired by the latest generation of probabilistic models for large-scale simulation of renewable electricity generation \cite{ludkovski2022large} which provide the critical underpinning for a proper statistical framing of the CFE target problem. These stochastic models of the short term joint behavior of load and renewable production were modified to produce realistic long-term (multi-year) simulations. As a consequence the results here are fundamentally probabilistic in contrast to so-called ``8760" analysis of \emph{expected} volumes which by construction cannot yield distributional results and are, therefore, blunt instruments in transaction structuring. This simulation methodology has been used to structure 24/7 CFE transactions for large data center portfolios in PJM, NYISO and ISONE regions (see \cite{IronMountain}). 
 
The structures discussed here are useful for supporting the continued expansion of the renewable generation footprint, providing enhanced revenue streams to those with the foresight and investment appetite to build existing generation, while simultaneously incentivizing new builds with the prospect of enhanced profitability. Two white papers from Google \cite{GoogleManager, GoogleRoadmap} outline potential ways to scaling this market.

Due to the nonlinearity introduced by the capped ratio, probabilistic modeling (including the realistic representation of correlation within the CFE portfolio) is essential. 
We view the problem at an annual ($T=8760$ = 365 x 24 hours) scale, with hourly matching. The CFE score is taken as a constraint, with the objective of minimizing generation cost.

To handle the intrinsic stochasticity, we work with a collection of scenarios, jointly simulated across loads and generators. These scenarios represent potential fluctuations that render the realized CFE score a random variable. In line with this, we interpret the CFE target probabilistically, so that it must hold for a given (large) fraction of the scenarios.

Our contributions are two-fold: (1) we provide a mathematical framework of optimization with probabilistic constraints for CFE structuring, highlighting the role of the two primary parameters of CFE target  and guarantee level; (2) we present two realistic case studies where the underlying scenarios are fully probabilistically coupled and calibrated to a realistic renewable assets. As complement to this article, we provide an interactive online dashboard \cite{Shiny} where the users can themselves explore CFE portfolio characteristics.

The rest of the paper is organized as follows. Section \ref{sec:method} introduces our optimization problem; Section \ref{sec:optim} discusses our solution method; Sections \ref{sec:case-study}-\ref{sec:multiple-loads} present a single-load and two-load case study. Finally, Section \ref{sec:conclude} discusses future research.

\section{Methodology}\label{sec:method}
We consider a collection of renewable energy assets which we structure into a fixed portfolio with the goal of minimizing the respective energy procurement cost, while guaranteeing that a target CFE score is achieved with a given probability based on a finite set of scenarios. 

\begin{table}[h!]
\begin{tabular}{p{1.2cm}p{6cm}}
 & \textbf{Notation Used}  \\
$I$: & number of assets (energy sources), $i=1,\ldots, I$\\
$N$: & number of scenarios, indexed by $n=1,\ldots, N$\\
$T$: & number of time periods indexed by $t=1,\ldots, T$\\
$G_i(t)$: & amount of energy (MW) generated by source $i$ in hour $t$. $G_i^{(n)}(t)$ is the $n$-th scenario of $G_i(t)$\\
$L(t)$: & Load (MW) in hour $t$ with scenarios $L^{(n)}(t)$ \\
$p_C$ : & CFE target score, $p_C \in (0,1]$ \\
$\alpha$: & CFE quantile guarantee level, $\alpha \in (0, 1)$\\
$w_i$: & fraction of source $i$ procured for the portfolio\\
${\cal W} \subseteq \mathbb{R}^I$: & set of feasible portfolio allocations \\
$c_i$: & cost of energy ($\$/\text{MWh}$)  from source $i$\\
$\pi(t)$: & constructed CFE portfolio $\pi(t) = \sum_i w_i G_i(t)$ \\
\end{tabular}
\vspace*{-10pt}
\end{table}

Using the notation below, our problem is translated into building a static portfolio $\bw=(w_1, w_2, \ldots, w_I)$ 
representing an annual PPA that procures $\pi(t) = \sum_i w_i G_i(t)$ 
throughout the year. Let $\cR(t) := \min(1, \pi(t)/L(t))$ be the hourly CFE fraction, with the cap indicating that over-generation is not transferable across time. 
The PPA structure $\pi$ should cover most of the load across hours and achieve a 
 CFE score target of $p_C \%$ (e.g.~$p_C=0.95$, meaning that 95\% of the load must be matched with the CFE generators) with a probability of $\alpha\%$, 
meaning that the $1-\alpha$ quantile of the CFE score across all scenarios must be higher than $p_C$. Costs are assessed per MW of energy delivered, which is a random quantity.
Our goal is to minimize the expected PPA cost,  $\mathbb{E}\left[\sum_{i=1}^I w_i c_i\sum_{t=1}^{T}{G_{i}(t)}\right]$. This is equivalent to minimizing  the average hourly cost $\sum_{i=1}^I c_i \bar{g}_i w_i$ where $$\bar{g}_i := \frac{1}{NT}\sum_{n=1}^N\sum_{t=1}^TG^{(n)}_{i}(t) \simeq \mathbb{E}\left[ Ave_t \; G_i(t) \right]$$ represents the average generation for asset $i$ across all scenarios $N$ and time steps $T$. 

Note:  $\bw = (w_1, \ldots, w_I)$ the vector of the weights is a vector of size $I$, $\mathbf{G}$ is a $I \times N \times T$  3-d tensor, and $\mathbf{L}$ is a $N \times T$  matrix.

\section{The optimization framework}\label{sec:optim}

The mathematical formulation of the CFE structuring problem is as follows:

\begin{subequations}\label{eq:optimizationProblem_onlLoad} 
\begin{equation}
    \min_{\bw\in \mathcal{W}} \sum_{i=1}^I c_i \bar{g}_i  w_i 
\end{equation}
\begin{equation} 
\label{eq:P1-constraint}
    \text{s.t.} \; \text{quantile}_{1-\alpha} \left( \bcR(\bw) \right) \geq p_{C},
\end{equation}
\end{subequations}
where $\bcR(\bw) := \frac{1}{T} \sum_{t=1}^{T} \cR(t; \bw)$ is the random variable denoting the annual CFE score and 
 $\cR(t; \bw) := \min\left(\frac{\sum_{i=1}^I w_i G_i(t)}{L(t)}, 1\right)$ is the hourly CFE ratio,  reflecting the fraction of load matched, capped at unity.

\begin{remark}
    Our framework is similar to the portfolio management problem in which a manager maximizes expected returns with a quantile constraint guaranteeing that the portfolio return will be above a certain threshold with a given probability level. However unlike that analogue, we face a temporal dimension where surplus energy in one period cannot be utilized to compensate for energy deficits in other periods.
\end{remark}

Almost sure constraints $\alpha=1$ are straightforward to handle, since they correspond to $N$ separate piecewise-linear constraints $\bar{\cR}^{(n)} \ge p_C \forall n$. 
However, in the probabilistic setting with a continuum of potential scenarios, achieving the target in all possible outcomes is not meaningful. Practitioners have thus also considered the average-case constraint $\mathbb{E}[ \bcR ] \ge p_C$ which however does not give any control on the distribution of $\bcR$. Our goal is to span this gap by taking $\alpha \in (0.5, 1)$. In tandem, allowing a fraction of the scenarios to breach the target lowers the cost, providing another tuning knob for portfolio structuring.
Such probabilistic constraints are in general not solvable exactly; for example taking our prototypical case of $N=1000$ scenarios and $\alpha=0.95$, we require that the CFE target score $p_C$ would be achieved in at least 950 out of the 1000 scenarios. This in principle requires to search over all potential subsets of size 950 and finding  the minimal cost for each such subset to achieve $\bcR^{(n)} \ge p_C$.

Such ``robust'' optimization framework can be solved using mixed-integer programming (MIP)\cite{wolsey1999integer}. The idea would be to introduce binary variables to model whether a particular scenario meets the procurement constraint, translating to $\sum_{n=1}^N a_n \geq \alpha \times N$ for $N$ auxiliary binary variables,  $a_n := \mathbbm{1}_{\{\bar{\cR}^{(n)} \geq p_C\}}$. The introduction of a large number of auxiliary variables significantly increases the computational complexity, leading to very long solution times. The approach we propose considers the constraint in \eqref{eq:P1-constraint} directly as a black-box, non-convex function $\bar{\cR}(\bw)$ of the weight vector $\bw$. We observe that it is in fact piecewise linear from an hour-by-hour and scenario-by-scenario perspective, for a total of at most $T \times N$ pieces.
Arguing by the Law of Large Numbers that $\bcR(\bw)$ is approximately Gaussian, one can approximate 
 its $1-\alpha$ quantile by $\hat{\mu}(\mathbf{w}) + \Phi^{-1}(1-\alpha)\hat{\sigma}(\mathbf{w})$, where $\hat{\mu}(\mathbf{w})$ and $\hat{\sigma}(\mathbf{w})$ are the estimates of $$\mu(\mathbf{w}) = \mathbb{E}\left[\bcR(\bw) \right] \quad \text{and}\quad \sigma(\mathbf{w}) = \sqrt{\text{Var}\left( \bcR(\bw) \right)},$$
 and $\Phi$ is the standard Gaussian CDF.
 The problem is then reduced to:
\begin{equation}\label{eq:ApproxProblem}
    \begin{aligned}
    &\min_{\bw} f(\bw) \qquad f(\bw) := \sum_{i=1}^I c_i \bar{g}_i w_i \\
    \text{s.t.}& \qquad g(\bw) := \hat{\mu}(\mathbf{w}) + \Phi^{-1}(1-\alpha)\hat{\sigma}(\mathbf{w}) - p_C \geq 0.
\end{aligned}
\end{equation}
This problem can be linearized and solved by means of linear programming techniques. Following this linearization idea, we consider the constraint as a function of the weights $\mathbf{w}$, i.e.,
$$
g(\mathbf{w}) := \text{quantile}_{1-\alpha}\left(\frac{1}{T}\sum_{t=1}^T\min\left(\frac{\sum_{i=1}^I w_i G_i(t)}{L(t)}, 1\right)\right)
$$
and use the Sequential Least Squares Programming (SLSQP) algorithm to deal with the constrained minimization problem in \eqref{eq:ApproxProblem}
The SLSQP algorithm, first proposed by \cite{powell197031, fletcher1972algorithm}, is widely used to solve nonlinear least-squares problems. The algorithm works by iteratively solving a sequence of quadratic programming subproblems. At each iteration, a quadratic approximation of the Lagrangian function  $L(\bw, \lambda) = f(\bw) + \lambda^T g(\bw)$ is constructed around the current iterate. This quadratic approximation is then minimized subject to the constraints and is used to generate a new iterate. The SLSQP algorithm can be summarized as follows (see \cite{kraft1988software} for more details):

\begin{enumerate}
\itemindent=-9pt
    \item[1] Initialize the iterate $\bw_0$.
    \item[2] For $k = 1, 2, \ldots$
    \begin{enumerate}[align=right,itemindent=0em,labelsep=5pt,labelwidth=1em,leftmargin=0pt,nosep]
        \item[a]  Construct a quadratic approximation of the Lagrangian function around $\bw_{k}$, $L(\bw_k + \bh) \approx L(\bw_k) + \nabla L(\bw_k)^T \bh + \frac{1}{2}\bh^T \mathcal{H} (\bw_k) \bh$.
        \item[b] Solve the quadratic subproblem given by
        \begin{align*}
            & \min_{\bh} \frac{1}{2}\bh^T \mathcal{H}(\bw_k) \bh + \nabla L(\bw_k)^T \bh\\
            & \text{ s.t. } \quad \nabla g(\bw_k)^T \bh + g(\bw_k) \leq 0
        \end{align*}
        to obtain a search direction $\bh_k$.
        \item[c] Update the iterate: $\bw_{k+1} = \bw_k + \gamma_k\cdot \bh_k$ where $\gamma_k$ is a step size.
    \end{enumerate}
\end{enumerate}
In step 2.a, the gradient $\nabla L(\bw_k)$ and Hessian $\mathcal{H}(\bw_k)$ at $\bw_k$ are approximated via finite difference schemes. The algorithm stops whenever $f(\bw_k + \gamma_k\cdot \bh_k) > f(\bw_k)$ or the new iterate does not satisfy the constraints. The step size $\gamma_k$ is selected via a line search according to Wolfe's condition \cite{wolfe1969convergence}.
To address the potential convergence of the SLSQP algorithm to a local minimum of the objective function,
we initialize $\bw_0$ as the solution to the approximate problem  in \eqref{eq:ApproxProblem}.

\section{Illustrative Case Study I}\label{sec:case-study}

\begin{figure}
\centering
\includegraphics[width=0.45\textwidth,trim=0.2in 0.3in 0.2in 0in]{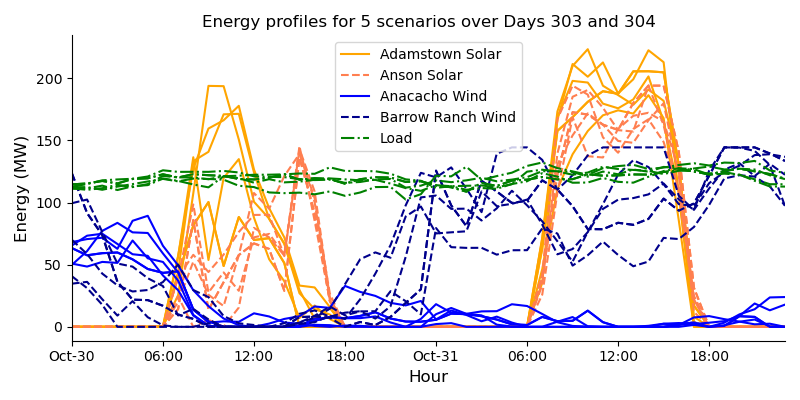}
\caption{Sample scenarios on two days (Oct 30-31: 303rd and 304th days of the year) for 2 solar assets, 2 wind assets and load. \label{fig:scenarios}}
\end{figure}

Our first case study features 2 solar assets (Adamstown and Anson), 2 wind assets (Anacacho and Barrow Ranch) and a hydro asset, taken to be fully deterministic. The underlying data and asset names refer to the ERCOT region and the ARPA-E PERFORM Dataset \cite{arpa-e}. Figure~\ref{fig:scenarios} shows 5 representative scenarios on 2 consecutive representative days for this collection of 4 renewable assets and the desired load. We emphasize that all simulations are fully coupled. The basic structure of high levels of solar CFE production during mid-day and with wind production predominantly at night shows that matching to a much less volatile load is quite challenging. 
Table \ref{tbl:corr} shows that the 2 Solar assets are highly correlated, while the 2 Wind assets are only moderately correlated. As is common, there is a positive dependence between Load and Solar, but a negative one between Load/Solar and Wind generation.  These features can be observed in Figure~\ref{fig:scenarios} where Barrow Ranch generation is high on Oct 31 (hitting its max capacity around midnight), while Anacacho is barely producing; in contrast there is a strong correlation between Adamstown and Anson. 

\begin{table}[!ht]
\caption{Hourly correlation among CFE sources of Case Study I. \label{tbl:corr}}
\centering {\small  \begin{tabular}{rrrrrr}
 & Adam. & Anson & Anac. & BaRa & Load \\ \hline 
Adamstown~Solar & 1.00 & 0.91 & -0.38 & -0.39 & 0.15 \\
Anson Solar & 0.91 & 1.00 & -0.40 & -0.40 & 0.15 \\
Anacacho Wind & -0.38 & -0.40 & 1.00 & 0.43 & -0.18 \\
BarrowRanch Wind & -0.39 & -0.40 & 0.43 & 1.00 & -0.22 \\
Load & 0.15 & 0.15 & -0.18 & -0.22 & 1.00 \\ \hline 
\end{tabular}}
\end{table}

\begin{table}[!ht]
\centering
\caption{Asset summary and portfolio solution for Case Study I. \label{tbl:5s}}
\begin{tabular}{lrrrr} \hline
Asset & Capacity & Ave Gen & Cost/MW & $w^*_i$ \\ \hline 
Adamstown Solar & 250.0 & 67.6 & 30 & 0.228 \\ 
Anson Solar & 200.0 &  61.6 & 30 & 0.419 \\
Anacacho Wind & 89.3 & 25.9 & 50 & 1.000 \\
Barrow Ranch Wind & 144.3 & 51.5 & 60 &  0.725 \\
Hydro & 123.7 &  97.8 & 102 & 0.383 \\ \hline
\end{tabular}
\vspace*{-10pt}
\end{table}

After optimizing for $p_C=0.9$ at $\alpha=0.95$-level, we get the portfolio weight vector $\bw^*$ listed in Table \ref{tbl:5s}. Note that the unit costs here were designed to bear some resemblance to the ERCOT marketplace; elsewhere solar RECs are substantially premium, with hydro the least expensive. Our discussion is agnostic to variations in the cost stack used. Observe that Anacacho is fully used, illustrating the common feature that some of the optimal weights bind to the constraints $w_i \in [0,1]$, due to either an asset being not competitive and not getting picked at all, or it being scarce and hence fully procured $w_i = 1$. The resulting procurement cost per MW of Load is $\sum_i w^*_i c_i \bar{g}_i/Ave_t (L_t) = 74.268$. This is significantly higher than the costs in Table~\ref{tbl:5s} due to overgeneration: the constructed least-cost portfolio $\pi(\cdot)$ on average procures 125.3\% of Load.

\begin{figure}[!ht]
\centering
\includegraphics[width=0.475\textwidth,trim=0.1in 0.0in 0.2in 0.3in]{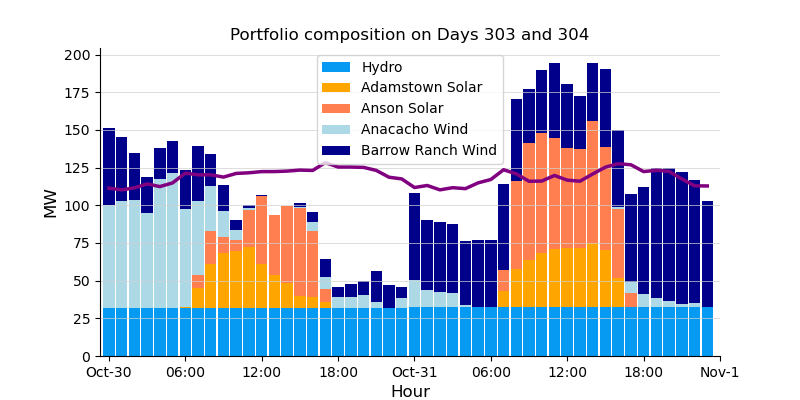}
\caption{Optimized CFE portfolio $\pi(t)$ on a representative scenario during two consecutive days. The asset universe includes 5 assets (2 Solar, 2 Wind, non-random Hydro). \label{fig:portfolio}}
\end{figure}

Figure \ref{fig:portfolio} shows the realized generation on the same 2 days based on the above $\bw^*$. We observe that in that specific scenario and time period, the constructed portfolio overproduces by up to 70\% (see 11am on Oct 31), however it has significant shortfall on the evening of the first day when there is no solar produciton and very little wind. 

\begin{figure}[!ht]
\centering
\includegraphics[width=0.475\textwidth,trim=0.5in 0.25in 0.65in 0.25in]{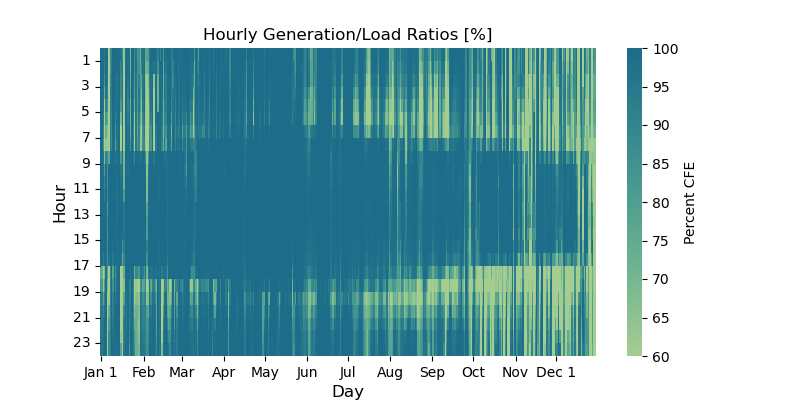}
\caption{Heatmap of the average hourly CFE score $\cR(t)$ across days and hours. The asset universe includes 5 assets (2 solar, 2 wind, non-random hydro). \label{fig:heatmap}}
\end{figure}

Figure \ref{fig:heatmap} displays the hourly average CFE scores $\mathbb{E}[ \cR(t; \bw^*)]$ across the entire year. By construction, the time-averaged CFE score is 90\% $=p_C$. The reliance on solar energy is clear, as the CFE score is close to 1 between 9-3pm and then is consistently below 0.9 in the late evening. 

Stronger CFE targets, i.e.~higher $p_C$ are more costly, and so is the higher level of guarantee $\alpha$. 
Table \ref{tbl:alpha-pc} shows how these 2 input variables drive the cost.
We observe that the guarantee level plays a mild role (less than 0.5\% additional cost) in portfolio cost, while  costs increase exponentially as CFE target $p_C \to 1$.  In the given setting, the maximum achievable average CFE target is 99.34\%, so that in most scenarios, even if procuring all available CFE sources, there are hours where load is not fully covered.

\begin{table}[!ht]
\caption{Portfolio cost (\$/MW of load) as a function of $\alpha$ (rows) and $p_C$ (columns) for Case Study I. \label{tbl:alpha-pc} }
\centering
\vspace*{-18pt}
$$\begin{array}{r|rrrrrrr}
 \alpha \backslash p_C    & 0.5 & 0.6 & 0.7 & 0.8 & 0.9 & 0.95 & 0.99 \\ \hline
0.5 & 19.38 & 25.60 & 33.53 & 46.09 & 73.07 & 92.55 & 126.26 \\
0.6 & 19.42 & 25.64 & 33.61 & 46.25 & 73.29 & 92.72 & 126.83 \\
0.7 & 19.46 & 25.71 & 33.69 & 46.41 & 73.47 & 92.88 & 127.41 \\
0.8 & 19.50 & 25.78 & 33.79 & 46.61 & 73.68 & 93.11 & 128.19 \\
0.9 & 19.56 & 25.87 & 33.92 & 46.87 & 74.01 & 93.40 & 129.13 \\
1.0 & 19.92 & 26.36 & 34.56 & 48.10 & 75.56 & 94.93 & 133.13 \\ \hline
\end{array}$$
\vspace*{-10pt}
\end{table}

\subsubsection{\textbf{Marginal Cost of Load}} Another way to explore the portfolio construction is to consider the \emph{marginal} portfolio, i.e.~which assets are used to match the next bit of additional load. This is captured by comparing the portfolios when an infinitesimal amount of additional load is added (multiplicatively, to reflect the non-constant temporal profile). In the case study, a marginal increase in load when $p_C=0.9$ is met with the marginal $\partial \bw/\partial L = [0.071, 0.215, 0, 0.421, 0.293]$ showing that at the margin it is mostly Anson Solar and BarrowRanch being used (Anacacho already being completely utilized, see Table \ref{tbl:5s}). In contrast, when $p_C = 0.95$ the marginal portfolio is $\partial \bw/\partial L = [0.071, 0.155, 0, 0.255, 0.519]$ shifting more into Hydro being the main marginal source of additional generation due to stricter CFE targets.

\subsubsection{\textbf{Effect of Diversification}} We augment the previous case study with additional renewable assets, for a total of 5 solar (fixed cost 30 \$/MWh) and 5 wind farms (50 \$/Mwh),
We then compare different combinations, i.e.~subsets of these 10 assets, comprising $n=1,\ldots, 5$ Solar, $n$ Wind, and $1$ Hydro asset, to investigate diversification impact, focusing on cost and downside risk. Our metric for evaluating downside risk is the value-at-risk (VaR) of the energy mismatch between load and  CFE generation over the entire year. 

We define the relative procurement shortfall $Y$, as the summation of the hourly shortfalls between load and constructed portfolio $\pi^*$: $$Y := \sum_{t=1}^T\min(\pi(t; \bw^*) - L(t), 0)/L(t).$$
For a level $\beta$, the VaR of the shortfall $Y$ is
$\text{VaR}_{\beta}(Y) = -\inf \{ y \in \mathbb{R} : \mathbb{P}(Y \leq y) \geq \beta \}$. Intuitively, $Y$ will be around $1-p_C$ the average fraction of load not met, and $\text{VaR}_{\beta}(Y)-p_C$ measures the downside risk of not meeting the CFE target.

\begin{table*}
\centering
\caption{Portfolios $\bw^{(k)}$ and optimized Costs $v_k$ for Matching $K=2$ loads with $I=3$ CFE sources in Case Study II.  Bolded values indicate minima of the respective column. \label{tbl:2loads} } 
{\vspace*{-16pt}} $$\begin{array}{lccccc} \hline 
 \text{Method} & \bw^{(1)} & v_1 & \bw^{(2)} & v_2 & v_1+v_2 \\  \hline 
\text{L1 priority} & [0.108, 0.159, 0.456] & \mathbf{45.139} &  [0.145, 0.304, 0.544] & 30.939 & 76.078 \\
\text{L2 priority} & [0.166, 0.231, 0.342] & 46.509 & [0.052, 0.268, 0.658] &  \mathbf{30.187} &  76.696 \\
\text{Split} &  [0.115, 0.150, 0.461] &  45.495 &  [0.169, 0.344, 0.500] &  31.586 &  77.081 \\
\text{Joint} & [0.144, 0.206, 0.380] &  45.774 &  [0.083, 0.274, 0.620] &  30.217 &  \mathbf{75.991} \\ \hline
\end{array}$$ \vspace*{-10pt}
\end{table*}

\begin{figure}[!ht]
\centering
\includegraphics[width=0.475\textwidth,trim=0.1in 0.3in 0.1in 0.3in]{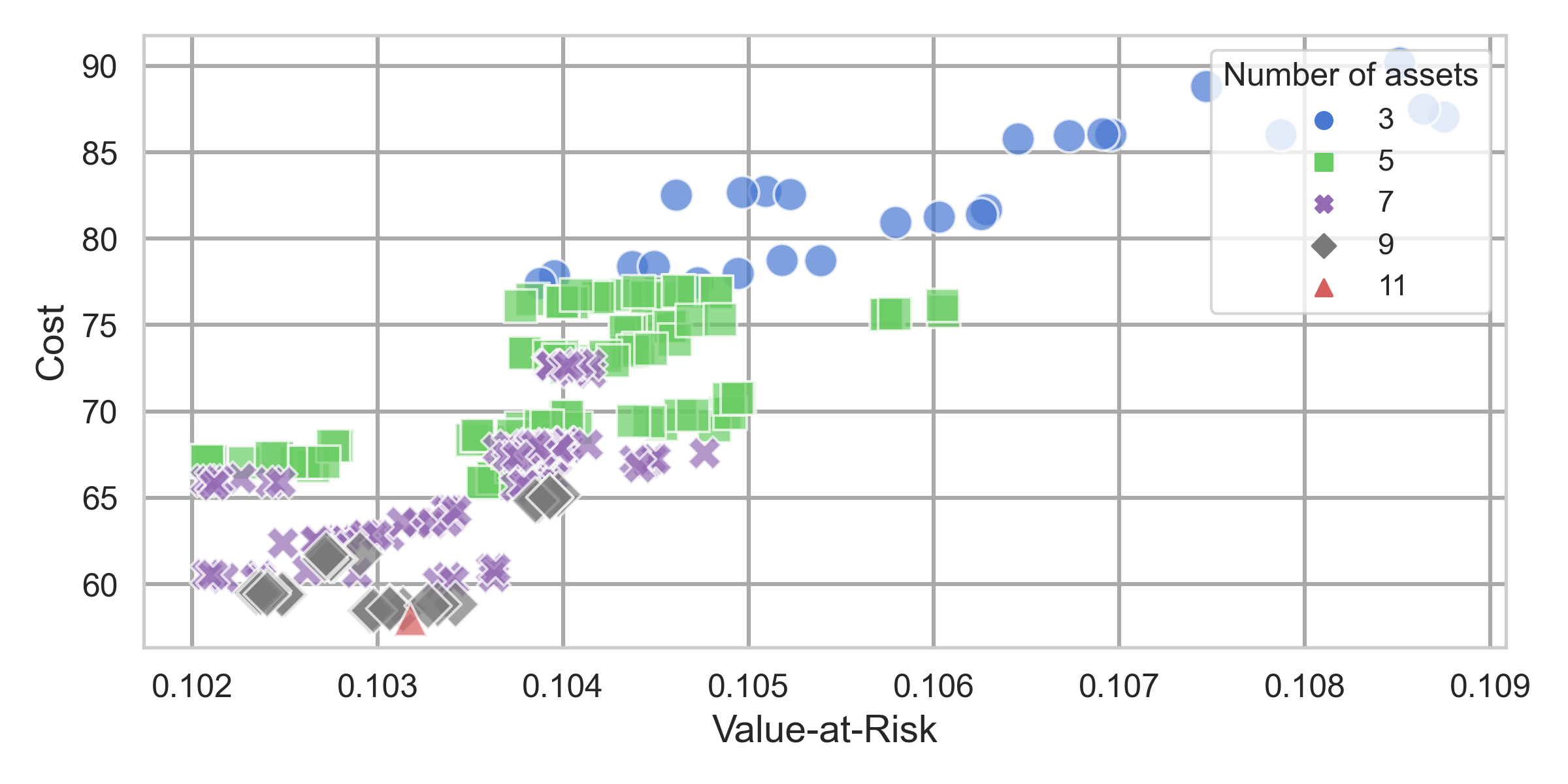}
\caption{Impact of diversification on the cost and 90\% CFE shortfall VaR. We evaluate optimal portfolios for all subsets of size $2n+1$ that consist of $n$ Solar and $n$ Wind assets, as well as the Hydro asset. \label{fig:diversification}}
\end{figure}

Figure \ref{fig:diversification} compares  various portfolio compositions 
 considering their respective costs and  CFE shortfall VaR at level $\beta=95\%$. As expected, the more energy sources are available, the lower the cost. We also see that through more diversified portfolios, the amount of mismatch between the demand and generation gets smaller for a similar cost.  
 
\section{Multiple Loads}\label{sec:multiple-loads}
In the multiple loads framework, our goal is to match heterogeneous load profiles with a portfolio of renewable energy generators. Suppose there are $K$ Loads, each with their own CFE target $p_C^{(k)}$ and guarantee level $\alpha^{(k)}, k=1,\ldots,K,$ and $I$ energy sources.
When there are multiple loads, cost minimization needs to be defined precisely as it depends on the economic setting; so far CFE contracts have been bilateral with no pooling or netting of multiple contracts.  
We consider three approaches depending on how the CFE structurer allocates the resources to different Loads. 

\subsubsection{\textbf{Sequential Optimization}}
In the first approach, the structurer successively addresses the optimization of each Load one at a time.  The first Load gets priority and can use any/all of the assets; remaining unclaimed generation then cascades down to become the input for the second optimization for Load 2, and so forth. 
Mathematically, we solve Problem  (P\ref{eq:optimizationProblem_onlLoad}) for Load $k$ with the constraint $ \bw\in \mathcal{W}^{(k)}$ where 
$\mathcal{W}^{(k)}=[0, \max(0, 1-\sum_{j=1}^{k-1}w^{(j)}_1)]\times  [0, \max(0, 1-\sum_{j=1}^{k-1}w^{(j)}_2)]\times \ldots \times [0, \max(0, 1-\sum_{j=1}^{l-1}w^{(j)}_I)]$, to address that the energy supply for Load $k$ has been reduced by the procurement of Loads 1 through $k-1$.  

\subsubsection{\textbf{Multipartite Optimization}} The CFE sources are split equally among the Loads. If there are two Loads, each one will have access to half the generation of each asset.  That is, there is no longer any sequencing and we solve Problem (P\ref{eq:optimizationProblem_onlLoad}) with allocation set $\mathcal{W} = [0, \sfrac{1}{K}]^I$, independently for each of $K$ Loads.

\subsubsection{\textbf{Concurrent Optimization}} The $K$ portfolios are constructed by jointly minimizing the sum of the costs:
\begin{align}\label{eq:optimizationProblem_2Loads_concurrent}
       & (\bw^{(1)}_{opt}, \bw^{(2)}_{opt}, \ldots) = \argmin_{(\bw^{(1)}, \ldots, \bw^{(K)})} \sum_{k=1}^{K}\sum_{i=1}^I c_i \bar{g}_i  w_i^{(k)}  \\ \notag
    &\text{s.t. } \sum_{k=1}^{K}{\bw^{(k)}} \in \mathcal{W}, \quad
     \text{quantile}_{1-\alpha^{(k)}} \left( \bcR^{(k)}(\bw^{(k))} \right) \geq p^{(k)}_C \  \forall k,
    \end{align}
where $\bcR^{(k)}$ is the CFE score of the $k$-th Load. The joint approach does not provide any preference to one Load over another, focusing on (cooperatively) achieving a minimal total cost while respecting individual Load targets/constraints.

Table \ref{tbl:2loads} illustrates the impact of these choices for Case Study II which features 3 stochastic energy sources (``Solar" with cost \$50/MWh, ``Wind" (\$28/MWh) and ``Hydro" (\$27/MWh)) and 2 Loads representative of Industrial and Commercial profiles. While Load1 is smaller than Load2, it has a higher target $p_C^{(1)}= 0.9 > p_C^{(2)}= 0.8$ with same $\alpha^{(1)}=\alpha^{(2)}=0.9$.

We observe that the Hydro asset is the most desirable among the 3 sources. Without constraints $L^{(1)}$ would take 45\% of it, while $L^{(2)}$ would take 66\% of Hydro, creating a competition for this asset, see the first two rows in Table~\ref{tbl:2loads}. If the assets are arbitrarily split down the middle, then this restriction does not bind for Load1, but hurts Load2 even more, as they can only get at most 50\% of Hydro (whereas they could get nearly 55\% as ``leftovers'' after Load1). Finally joint optimization is almost as cheap for Load2 as being first in line, and only hurts Load1 a bit, leading to the lowest sum of procurement costs (right-most column).

\section{Conclusion}\label{sec:conclude}

We have presented initial analysis of structuring portfolios of renewable assets to probabilistically achieve given CFE targets at minimal cost. Our work provides a methodological underpinning to the recent bilateral structures that have been announced in the industry, such as the previously referenced Iron Mountain transaction structured by RPD Energy, as well as the Google data center commitments. Looking ahead, two avenues of research will be explored in follow-up projects. First, concurrently handling multiple loads (e.g.~ for $K \in [5,10]$) presents a methodological challenge; one seeks a structure that combines features of the sequential, priority-driven allocation (as some Loads might have some sort of preferred status), and the joint cooperative solution that minimizes total procurement costs. Second, one should more explicitly define non-compliance costs, i.e. sustaining a realized CFE score below the target. In practice, the provider might be on the hook for some costs, e.g. to procure in real-time additional CFE credits in order to reach the target when the portfolio is performing below expectations. Properly capturing this cost of shortfall would modify the minimization objective. 

\bibliographystyle{IEEEtran}
\bibliography{pesBib}

\end{document}